\documentclass[amsmath,amssymb,aps]{revtex4-2}

\usepackage{caption}

\usepackage{subcaption}
\usepackage{graphicx}

\usepackage{dcolumn}
\usepackage{bm}
\usepackage{braket} 
\usepackage{hyperref}  

\hypersetup{
    colorlinks = true
}

\newtheorem{defn}{Definition}
\newtheorem{definition}[defn]{Definition}

\mathchardef\mhyphen="2D

\begin{document}

\preprint{APS/123-QED}

\title{Benchmark of the Full and Reduced Effective Resistance Kernel for Molecular Classification}

\author{Adam Wesołowski}
\email{adam@qunasys.com}
\author{Karim Essafi}
\email{karim@qunasys.com}
\affiliation{QunaSys, Ole Maaløes Vej 3, DK-2200, Copenhagen, Denmark}

\date{\today}

\begin{abstract}
We present a comprehensive study of the commute time kernel method via the effective resistance framework analyzing the quantum complexity of the originally classical approach. Our study reveals that while there is a trade-off between accuracy and computational complexity, significant improvements can be achieved in terms of runtime efficiency without substantially compromising on precision. Our investigation highlights a notable quantum speedup in calculating the kernel, which offers a quadratic improvement in time complexity over classical approaches in certain instances. In addition, we introduce methodical improvements over the original work on the commute time kernel and provide empirical evidence suggesting the potential reduction of kernel queries without significant impact on result accuracy. Benchmarking our method on several chemistry-based datasets: $\tt{AIDS}$, $\tt{NCL1}$, $\tt{PTC-MR}$, $\tt{MUTAG}$, $\tt{PROTEINS}$ - data points previously unexplored in existing literature, shows that while not always the most accurate, it excels in time efficiency. This makes it a compelling alternative for applications where computational speed is crucial. Our results highlight the balance between accuracy, computational complexity, and speedup offered by quantum computing, promoting further research into efficient algorithms for kernel methods and their applications in chemistry-based datasets.  
\end{abstract}

\maketitle


\section{\label{sec:level1}Introduction}

Kernel methods are a cornerstone of modern machine learning and statistical analysis, offering powerful tools for pattern recognition, classification, and regression tasks (see \cite{kriege2020survey, nikolentzos2021graph} for a comprehensive review on the theory for graph kernels). By implicitly mapping data into high-dimensional feature spaces, kernel methods can reveal linear relationships in otherwise complex nonlinear data inherent in many real-world datasets. A key advantage of kernel methods lies in their ability to operate in these high-dimensional spaces without explicitly computing the transformations, leveraging the so-called "kernel trick" to perform computations directly in the original input space.

Among their myriad of applications, kernel methods have shown to be very promising in the field of chemistry, where they play a crucial role in a variety of tasks, including molecular property prediction\cite{schapin2023}, quantitative structure-activity relationship (QSAR) modeling\cite{mauri2017}, and materials discovery. By effectively capturing the underlying patterns in chemical data, kernel methods facilitate the prediction of molecular behaviors, aiding in the design of new compounds with desired properties.

Recent advances have seen the development of a new kernel method based on the concept of effective resistance, a measure derived from electrical network theory. Effective resistance provides a unique way of quantifying the similarity between data points by considering the connectivity and interactions within a network. In the context of chemical data, this approach can offer insightful perspectives on molecular structures and their interactions, potentially leading to more accurate and interpretable models.

The commute time kernel, introduced in 2010\cite{commutetimekernel}, was an early attempt to leverage graph structures for clustering tasks. Ever since, numerous methods have been proposed, most notably the deep graph convolutional neural network(DGCNN) \cite{10.5555/3504035.3504579}, which currently constitutes the state-of-the-art. In light of advancements in classical and quantum algorithms for the effective resistance estimation\cite{spielman2008graph, apers2022elfs}, we revisit the theory of a kernel based on the effective resistance. As shown by \cite{commtime=effres} the effective resistance is equal to the commute time re-scaled by the size of the graph. 

Recently a quantum algorithm for the estimation of effective resistance have been proposed\cite{apers2022elfs}. This has opened a venue to reconsider the commute time kernel as an effective resistance kernel due to the possibility of decreased time complexity of computing the kernel. We further refine the method by showing that it is not necessary to compute the kernel between all pairs of nodes in a graph, as originally required \cite{commutetimekernel}, to obtain satisfactory accuracy. Moreover, using recently proposed quantum algorithms, we show substantial reductions in the time complexity of the commute time/effective resistance kernel.

Finally, we provide a benchmark of the method on standard chemistry datasets - $\tt{AIDS}$, $\tt{NCL1}$, $\tt{PTC-MR}$, $\tt{MUTAG}$, $\tt{PROTEINS}$ -  which were missing from the literature, as well as on the discrimination of Erdős–Rényi random graphs of the same size with slightly varying parameter $p$. Our results are compared to both classical kernel methods and the state-of-the-art approach of DGCNN, positioning our approach as a promising alternative where computational efficiency is paramount.

In section II we outline the background knowledge, section III contains our definitions of the kernels, and section IV contains the results of the conducted benchmarking. We conclude with a discussion and directions for further work in section V.

\begin{figure}
    \centering
    \includegraphics[width=0.65\linewidth]{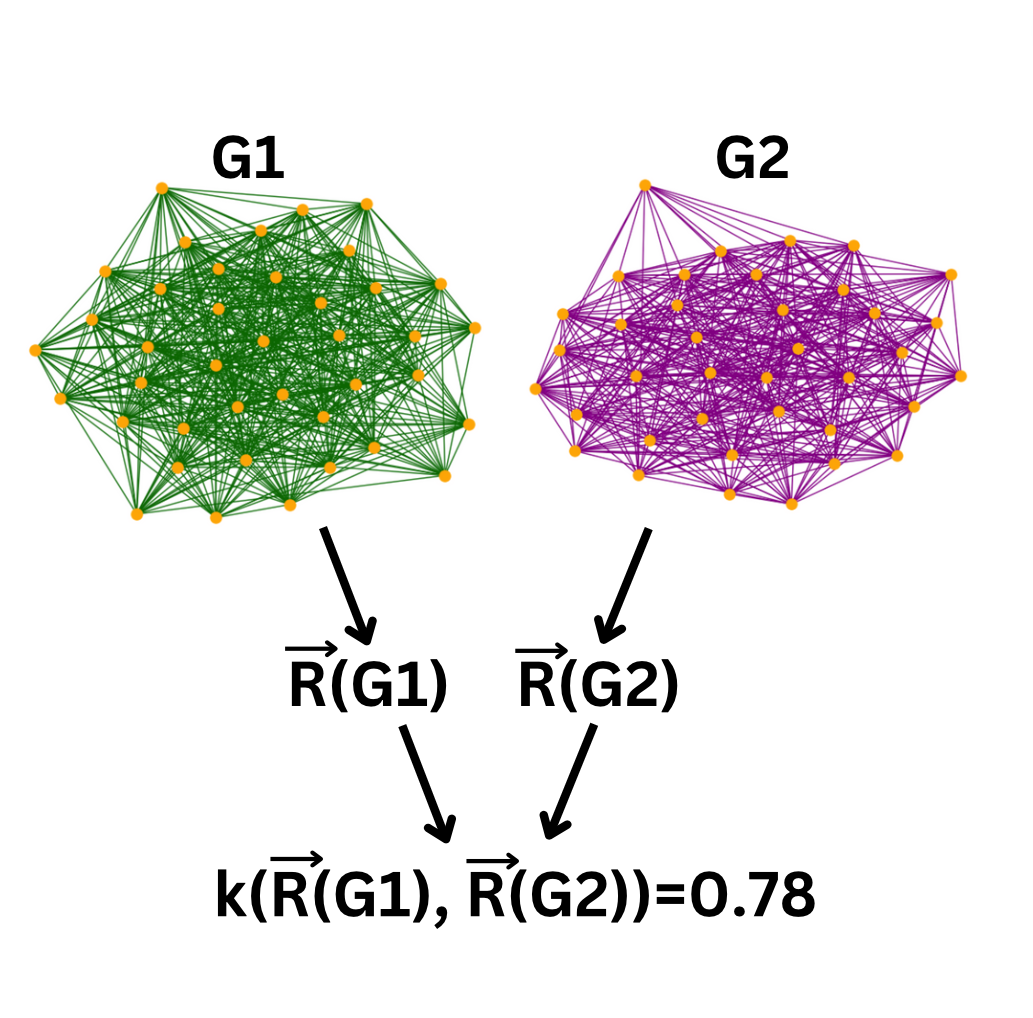}
    \caption{Kernel $k$ of two graphs $G_1$ and $G_2$}
    \label{fig:graph_kernel}
\end{figure}


\section{Background}
\subsection{Graph kernels}
A kernel $k$ is a positive semi definite function that given a pair of feature vectors of graphs $G_1$ and $G_2$  assigns a real-value number to them $G_1 \times G_2 \rightarrow \mathbb{R}$ (see Figure  \ref{fig:graph_kernel}). 
\begin{definition}
    A \textit{Mercer kernel} is a function that defines a valid kernel, which corresponds to an inner product in some (potentially high-dimensional) feature space. Formally, let $X$ be a non-empty set. A function $k: X \times X \to \mathbb{R}$ is called a Mercer kernel if it satisfies the following conditions:

\begin{itemize}
    \item \textbf{Symmetry:} For all $x, y \in X$, the kernel function $k$ must satisfy:
    \begin{align}
    k(x, y) = k(y, x)
    \end{align}
    \item \textbf{Positive Semi-Definiteness:} For any finite set of points $\{x_1, x_2, \ldots, x_n\} \subseteq X$ and any real numbers $c_1, c_2, \ldots, c_n$, the kernel function $k$ must satisfy:
    \begin{align}
    \sum_{i=1}^{n} \sum_{j=1}^{n} c_i c_j k(x_i, x_j) \geq 0
    \end{align}
\end{itemize}
 That is, there exists a mapping $\phi: X \to \mathcal{H}$ such that:
\begin{align}
k(x, y) = \langle \phi(x), \phi(y) \rangle_{\mathcal{H}}
\end{align}
for all $x, y \in X$, where $\langle \cdot, \cdot \rangle_{\mathcal{H}}$ denotes the inner product in a Hilbert space $\mathcal{H}$.

\end{definition}
The value of Mercer kernels lies in their suitability for use in data classification algorithms, in particular the support vector machine (SVM).
In this work, the commute time kernel is defined as a radial basis function (RBF), which is known to be a Mercer kernel. Thereby, the commute time/effective resistance kernel also is a Mercer kernel and thus can be used with SVMs.
\begin{definition}
    The Gram matrix, which encodes the pairwise similarities between graphs based on a chosen kernel function, is defined as follows:
\begin{align}
\mathbf{K} = \left[ k(G_i, G_j) \right]_{i,j=1}^n
\end{align}
where $ K_{ij} = k(G_i, G_j), \quad \forall \, i,j \in \{1, \dots, n\}$.
\end{definition}

\subsection{Graph theory}
Throughout the paper, we use technical terminology inherent to graph theory, which for clarity we outline in this subsection. 

\begin{definition}[Erdős–Rényi graphs
]An \textit{Erdős–Rényi random graph} $G(n, p)$ is a random graph with $n$ vertices where each pair of distinct vertices is connected by an edge independently with probability $p$.

Formally, let $V$ be a set of $n$ vertices, i.e., $V = \{v_1, v_2, \ldots, v_n\}$. For each pair of distinct vertices $(v_i, v_j)$ with $1 \leq i < j \leq n$, an edge $(v_i, v_j)$ is included in the graph with probability $p$. The inclusion or exclusion of different edges is independent of each other.
    
\end{definition}
\begin{definition}[Graph Laplacian]
    Given an undirected graph $G = (V, E)$ with vertex set $V = \{v_1, v_2, \ldots, v_n\}$ and edge set $E$, the \textit{Graph Laplacian} (or Laplacian matrix) $L$ is defined as follows: 
    
    Let $A$ be the adjacency matrix of $G$, where $A_{ij} = 1$ if there is an edge between vertices $v_i$ and $v_j$, and $A_{ij} = 0$ otherwise. Let $D$ be the degree matrix of $G$, which is a diagonal matrix where $D_{ii}$ is the degree of vertex $v_i$, i.e., the number of edges incident to $v_i$.
    The Laplacian matrix $L$ is given by:
\begin{align}
L = D - A
\end{align}
\end{definition}

More explicitly, the entries of the Laplacian matrix $L$ are defined as:
\begin{align}
L_{ij} = 
\begin{cases} 
\deg(v_i) & \text{if } i = j \\
-1 & \text{if } i \neq j \text{ and } (v_i, v_j) \in E \\
0 & \text{otherwise}
\end{cases}
\end{align}

where $\deg(v_i)$ denotes the degree of vertex $v_i$.

\subsection{Effective resistance}
In this section we briefly introduce crucial concepts that we make use of in this work. For a comprehensive study of the electric network formalism we refer the reader to chapters 2 and 9 in \cite{bollobas1998modern}. 
A graph $G$ can be viewed as an electrical circuit with two special vertices $s$ and $t$ called, a source and a sink to which the potential difference is applied. An edge between some vertices $a$ and $b$ is treated as a wire with resistance $R(a,b)$ and a potential difference $V_{a-b}=V_a-V_b$.
To introduce the concept of effective resistance between two nodes in a graph, we firstly need to introduce the concept of Moore Penrose pseudoinverse of a matrix. Eigenvalues of the Laplacian matrix $L$ are real, non-negative numbers $\lambda_i$ $i\in[0,N-1]$, with smallest eigenvalue $\lambda_0=0$. It follows that $L$ is singular, and thus the inverse matrix $L^{-1}$ does not exist. 
However, one can define the Moore-Penrose pseudoinverse $L^+$ of matrix $L$ as:
\begin{definition}
    Moore-Penrose pseudoinverse of matrix M is defined as a matrix $M^+$ that satisfies the following conditions:
    \begin{enumerate}
        \item $MM^+M=M$
        \item $M^+MM^+=M^+$
        \item $(MM^+)^{*}=MM^+ $
        \item $(M^+M)^{*}=M^+M$
    \end{enumerate}
These conditions ensure that the pseudoinverse exists and is unique.
Then the Moore-Penrose pseudoinverse $M^+$ of a matrix $M$ is defined such that:
\begin{align}
M^+ = V \Sigma^+ U^T
\end{align}
where $M = U \Sigma V^T$ is the singular value decomposition (SVD) of $M$, and $\Sigma^+$ is obtained by taking the reciprocal of each non-zero element on the diagonal of $\Sigma$, and then transposing the resulting matrix. 
\end{definition}

From the Laplacian $L$ of $G$ one can calculate the effective resistance between any two vertices in $G$ as:
\begin{align}
    R(a,b)=(e_a-e_b)^TL^{+}(e_a-e_b)=L_{aa}^{+}-2L_{ab}^{+}+L_{bb}^{+}
\end{align}
where $L^{+}$ is the Moore-Penrose pseudo-inverse of the singular matrix $L$. The $e_i$ is a vector with '0' entries everywhere except in position $i$ where it has value 1. 
The double subscripts indicate the position of the entry in the $L^{+}$ matrix.
Effective resistance, has many desirable properties, importantly it defines a \textit{metric} on a graph\cite{Klein1993ResistanceD,effresmetric}. In that regard, it is similar to a simple edge distance, based on which the shortest path kernel has been developed. Nevertheless, the notion of effective resistance between two nodes differs from the notion of a shortest path between the two nodes. The effective resistance usually captures the global graph structure between the two nodes, whereas the shortest path represents a more local information. Thus, in a general setting a single value of effective resistance between two vertices contains more information about the graph than the shortest path between the vertices.
Effective resistance is well known to have applications in graph characterization \cite{ZHOU2008120}. One of the popular related concepts is the total effective resistance, known also as the Kirchoff Index, defined as a sum of all pairwise effective resistances. The total effective resistance is deemed to be a good measure of robustness of the underlying graph network\cite{robustness}.
We calculate the value of effective resistance between nodes in a graph to obtain a feature vector of the original graph. We elaborate on the details of the kernel construction in section III.  
\subsection{Quantum versus classical computation of effective resistance
}
Classically, the computation of effective resistance has been well studied. For expander graphs the algorithms run very efficiently, however, it has been shown that for a general graph instance even with bounded degree the best classical algorithm runs in $\Omega(m)$ time\cite{spielman2008graph}, where $m$ denotes the number of edges. 

A quantum walk-based algorithm has been proposed by authors in \cite{apers2022elfs}. The algorithm estimates the effective resistance between two nodes in a graph, using $O\left(\frac{\sqrt{R(v,u)m}}{\varepsilon^{\frac{3}{2}}}\right)$ quantum walk steps per single pair estimation. The algorithm outputs an $\epsilon\mhyphen approximation$ of the effective resistance between a pair of vertices $v,u$. To avoid defining the time complexity of estimating the effective resistance with its value, one can always replace $R_G(v,u)$ with the length $l$ of the shortest path between $v$ and $u$ in $G$, as it always upper-bounds the values of $R_G(v,u)$. If one decides to calculate the full kernel, i.e between all $n(n-1)/2$ pairs of vertices, then classically the most efficient exact algorithm allows to do that in $O(n^3)$ steps, via explicit calculation of the Moore-Penrose pseudoinverse. Whereas using a quantum effective resistance estimation algorithm\cite{apers2022elfs} the time complexity, of evaluating all-pairs effective resistance can be bounded by $O(n^2\sqrt{R_{max}(G)m})$. Note that $R_{max}(G)$ is always bounded by the diameter $D$ of $G$, which in extreme cases can be as large as $n$. However for some natural classes of finite graphs it is usually bounded by some smaller function of $n$.  Whenever $D(G) = O(polylog(n))$ the quantum approach requires $O(n^2\sqrt{Dm})$ steps. 
In this work we propose to reduce the number of kernel evaluations from $O(n^2)$ to $O(\sqrt{n})$. In such case best classical approximation algorithm can compute the effective resistances for a single graph in time $O(\sqrt{n}m)$, whereas quantumly the time complexity scales as $\tilde{O}(n\sqrt{m})$. 
In the proposed version of the kernel of the graph the quantum time complexity for dense graphs is $\tilde{O}(n^{2})$, and for sparse graphs achieves the time complexity of $\tilde{O}(n^{1.5})$.
The preceding discussion pertains to the evaluation of kernels for individual graphs. The computational complexity of a single kernel evaluation represents the complexity of the given graph kernel. However, when employing a kernel with classical Support Vector Machines (SVM), for a dataset of size 
$N$, one must perform $N^2$ kernel evaluations to compute the kernel value between every pair of graphs. In established practices such as those discussed in \cite{nikolentzos2021graph, kriege2020survey} the $N^2$ multiplicative factor is typically omitted when detailing the complexity of the kernel, as this factor 
appears across all graph kernels.

In subsection A we explain the theoretical evidence that this approach should not lead to much decreased accuracy, which is then supported by modest empirical evidence we give in figures.(1-6). We note that as in many machine learning techniques it is hard to demonstrate any rigorous proofs of accuracy guarantees.

\section{Effective resistance graph kernel}

We note that a very similar approach has been proposed in \cite{commutetimekernel}, and we give a more comprehensive study and benchmarking results. In this work, we give both, with the focus on benchmarking the results on the standard chemistry datasets taken from the TU Dortmund online repository\cite{KKMMN2016}.

Given a graph $G(N,M)$, one calculates the all-pairs effective resistance, that yields a list of $\frac{n(n-1)}{2}$ numbers. Choosing this list as a representation of $G$ and attempting to compare two graphs, {\it i.e}  calculate a kernel between $G_1$ and $G_2$, leads to ambiguity. Given an operation on the lists, one may permute the values in lists, resulting possibly in different values for the kernel. To alleviate this ambiguity, we sort the lists in an increasing (or decreasing) order. This will allow us to compare the structure of graphs encoded in the values of effective resistances in an unambiguous manner. 
Given a graph $G$, one calculates effective resistance values between all pairs of vertices and constructs a vector $\vec{v}(G)$ whose entries are effective resistance values between vertices in G. The vector lives in the $n(n-1)/2$-dimensional Hilbert space $\mathcal{H}$, spanned by $n(n-1)/2$ basis vectors. Formally,
\begin{definition}
    Given a graph G(N, M), we associate with graph G a feature vector $v_G\in \mathbb{R}^{\frac{n^2-n}{2}}$ such that $v_G=(r_1,r_2,...,r_{\frac{n^2-n}{2}})$, where $r_i$ $i\in [1,2,...,\frac{n^2-n}{2}]$ are the values of the effective resistances between pairs of nodes in G and $r_1\leq r_2\leq ...\leq r_{\frac{n^2-n}{2}}$. 
\end{definition}
The vectors $v$ are real valued and are embedded in the space $\mathbb{R}^d$. The Euclidean distance between vectors $v$ constitutes a natural measure of dissimilarity of the underlying graphs. One can see that for isomorphic graphs the Euclidean distance between feature vectors is zero, and intuitively it grows indefinitely as graphs become less and less alike. Any $p\mhyphen norm$ of the difference of two feature vectors, constitutes a good quality measure for clustering of graphs, but does not satisfy requirements to be a valid kernel function. Nevertheless, a small modification of the distance function leads naturally to a Mercer kernel, that provably satisfies positive semi-definiteness condition and thereby constitutes a valid graph kernel.  
The effective resistance kernel between two graphs $G_1$ and $G_2$ with the corresponding vectors $v_{G_1}$ and $v_{G_2}$ is defined as follows:
\begin{definition}
    The effective resistance kernel $k(G_1,G_2)$ between two graphs $G_1$ and $G_2$, is defined through feature vectors of these graphs $\vec{v}_{G1}$ and $\vec{v}_{G2}$. 
    Vector entries correspond to effective resistances between (all) pairs of vertices in those graphs as per Def.2. The kernel k is then defined as:
    \begin{align}
        k(G_1,G_2)= e^{-\gamma||\vec{v}_{G_1}-\vec{v}_{G_2}||^{2}}
    \end{align}
\end{definition}  
Where $||\vec{v}||^{2}$ is a squared Euclidean norm of a vector $\vec{v}$.
When the number of vertices in graph $G_1$ is not equal to the number of vertices in graph $G_2$, then the corresponding vectors
$v_{G_1}$ and $v_{G_2}$ span Hilbert spaces of different dimensions and thus cannot be reliably compared using the kernel from Def.7. To overcome this technical issue, whenever dimensions of the vectors are not equal we add zeroes to the vector of smaller dimension, so that dimensions are equal. This leads to a problem of where to put the zeroes in the vector of smaller dimension. Suppose $dim(v_{G_1})< dim(v_{G_2})$ then different allocations of zeroes to $v_{G_1}$ will possibly lead to different values of the kernel in def.7. The question that arises, is which effective resistances are more important to compare, the small or the large ones? If one decides to add the zeros on the right-hand side of $v_1$, then one reliably compares the small values, assuming the entries are sorted in increasing order as per Def.6. If on the other hand one decides to add all zeroes from the left hand side then one reliably compares the largest values. Of course one can add zeroes in different places at will as long as the dimensions of vectors will be matched.
In our calculations, we make a choice of adding the zeroes from the RHS, such that the modified vector corresponding to the smaller graph has entries sorted in increasing order up to some point after which zero entries occur, and the number of zero entries is equal to the difference in dimensions between the two vectors. This simple modification assures that the kernel function in equation (2) is well defined.

When the task of evaluating the vectors for all graphs is complete, one can proceed to calculate kernels between all graphs resulting in a Gram matrix. The diagonal entries are equal to 1 which is a value that the effective resistance kernel assumes for isomorphic graphs. Whenever the two graphs are not isomorphic because of a different number of edges, the kernel is clearly greater than one, but the accuracy to which the effective resistance estimation needs to be performed to resolve the difference may be arbitrarily high. When there exist more subtle differences such as varying connectivity, the value of the kernel is also greater than 1, which follows from the fact that effective resistance is a metric. It may seem that the expressivity of the effective resistance kernel is substantial. However, it has a number of shortcomings. Firstly, it does not recognise structures. Be it trees or cycles, the effective resistance does not assume unique values for such structures. Secondly, the kernel between two non-isomorphic graph may be $\epsilon\mhyphen close$ to 1, requiring exponential accuracy of effective resistance estimation. This assures that the problem of computing the effective resistance kernel is \textit{Graph Isomorphism-complete} (in short \textit{GI-complete}). Proving the above, simultaneously shows that the kernel is not \textit{complete}, despite intuitive inclination to ascribe it a high expressibility.

\subsection{$\sqrt{n}$ evaluations is statistically sufficient}
In a graph $G$ with $n$ vertices it is enough to get $O\left(\frac{\sqrt{n}}{\epsilon^2}\right)$ distance samples $d(v_i,v_j)$ to estimate the average edge-distance between all $n(n-1)/2$ pairs of nodes to $(1+\epsilon)$ multiplicative accuracy\cite{resultsqrtn} (Theorem 5.1). That indicates that $O \left(\frac{\sqrt{n}}{\epsilon^2} \right)$ distance samples is statistically meaningful and perhaps should suffice to capture some global feature of the graph. Inspired by that finding, we put forward a conjecture that a kernel based on a different distance measure- the effective resistance, may require not the full $O(n^2)$ resistance distance evaluations but only $O(\sqrt{n})$, and still yield good accuracy. 
\begin{definition}[Reduced effective resistance kernel]
 The reduced effective resistance kernel $k(G_1,G_2)$ between two graphs $G_1$ and $G_2$, is defined through feature vectors of these graphs $\vec{v}_{G1}$ and $\vec{v}_{G2}$. 
    Vector entries correspond to effective resistances between $\sqrt{n}$ random pairs of vertices in those graphs as per Def.2. The kernel $k$ is then defined as:
    \begin{align}
        k(G_1,G_2)= e^{-\gamma||\vec{v}_{G_1}-\vec{v}_{G_2}||^{2}}
    \end{align}
    
\end{definition}
 In the reduced kernel we also deal we the aforementioned 'padding with 0's' problem. This is because the number of pairs $\sqrt{n}$ in a graph $G_1$ may not be equal to $\sqrt{n}$ in some other graph $G_2$, as previously we add zeroes to the vectors from the right hand side. Furthermore, We set $\gamma=1$.
\section{Results}
The features of a good kernel method are substantial expressibility, {\it i.e} ability to tell apart graphs, high accuracy and low time complexity of kernel calculation. In this section, we present benchmark results of our kernel. A testament to the expressibility of our kernel is a fact that it can correctly cluster random graphs of the same size with parameter $p$ varying on the scale of $\delta p=0.01$. Furthermore, we benchmark our method on standard chemistry datasets and show that its performance is either comparable or superior to similar kernel methods. For the experiments on the chemistry datasets, the results were obtained using SVM with the training parameter $C$ varying in the set $\{ 10^{-3},10^{-2},10^{-1},1,10^{2},10^{3}\}$ and the reported result corresponding to $C$ yielding best accuracy. For the SVM, classification datasets were split into 80\% training and 20\%test subsets, repeated 10 times on varying subsets such that accuracy reported is the mean value with standard deviation. Chemistry datasets were taken from TU Dortmund repository\cite{KKMMN2016}. Erdős–Rényi random graphs were generated using networkx Python library\cite{networkx}. 

\subsection{Discrimination of Erdős–Rényi random graphs.}
In this section, we provide a benchmark that was inspired in our case by the work of \cite{henry2021quantum}. The benchmark relies on the kernel's ability to discern Erdős–Rényi random graphs of the same size, with varying parameters $p$. The parameter is the probability that a random pair of nodes in the graph are connected. It can be thought of as the expected edge density of the graph. We think of $p$ as a $graph property$. Thus, the problem can be phrased as a decision problem: Does the given Erdős–Rényi random graph have $p$ equal to some value? Clustering of graphs according to their $p$ value becomes then the problem of interest. Figure 3 shows that the reduced effective resistance kernel is able to discern graphs where the difference of parameter $p$ is as small as $\delta p=0.01$. In the figures below we give a benchmark on 8 random graphs with 200, 400  and 600 nodes and varying parameter $p$, for clarity the heat map was added together with the numerical values. If the graphs are isomorphic then the value of the kernel is 1, in alignment with the definitions 8 and 9. By construction it follows that the larger the value of the kernel between two graphs the more 'similar' the graphs are. In all of the figures below graphs $G_1, G_2, G_3, G_4$ belong to one class, i.e have the same value of $p=p_1$, and graphs $G_5, G_6, G_7, G_8$ belong to another class with a different value of $p=p_2$, the specifications are given locally in the captions.

\begin{widetext}

\begin{figure}[htbp]
    \centering
    \subfloat[Graphs with 200 nodes. Graphs $G_1$--$G_4$ correspond to $p_1=0.85$, and graphs $G_5$--$G_8$ have $p_2=0.9$. Reduced effective resistance kernel was used for the calculation.]{\includegraphics[width=0.45\textwidth]{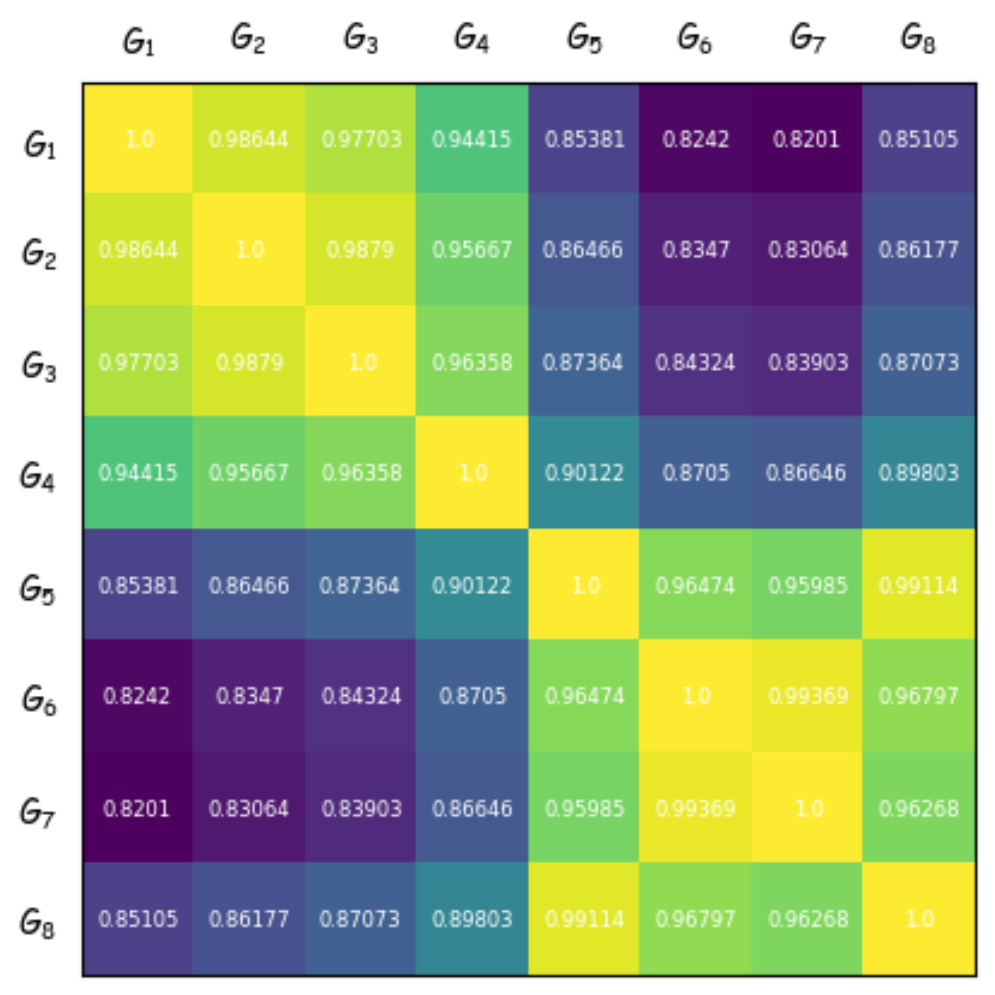} \label{fig:first}}
    \hfill
    \subfloat[Graphs with 200 nodes. Graphs $G_1$--$G_4$ correspond to $p_1=0.85$, and graphs $G_5$--$G_8$ have $p_2=0.9$. The original-all pairs effective resistance kernel was used for the calculation.]{\includegraphics[width=0.45\textwidth]{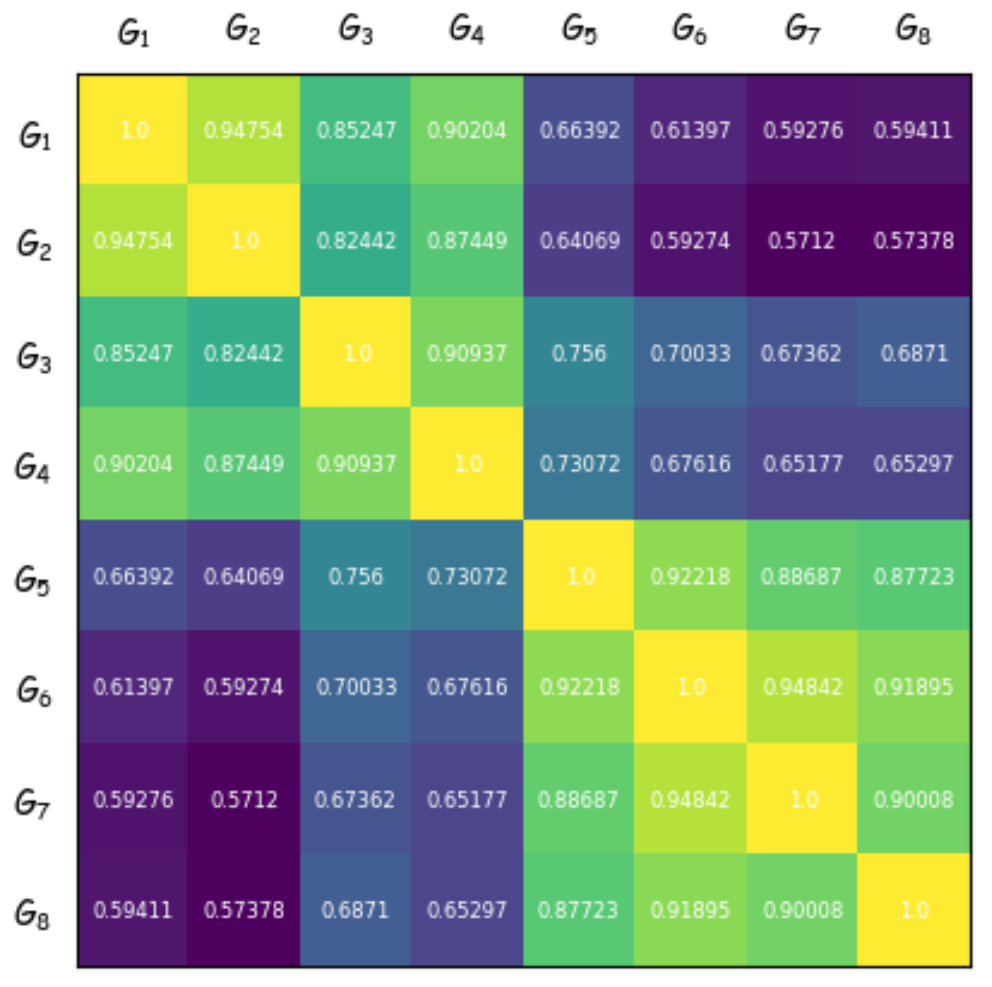} \label{fig:second}}
    \caption{Comparison of Gram matrices with kernel values of a) the reduced effective resistance kernel b) all-pairs effective resistance kernel.}
    \label{fig:comparison1}

\end{figure}

\begin{figure}[htbp]
    \centering
    \subfloat[Graphs with 400 nodes. Graphs $G_1$--$G_4$ correspond to $p_1=0.9$, and graphs $G_5$--$G_8$ have $p_2=0.92$. Reduced effective resistance kernel was used for the calculation.]{\includegraphics[width=0.45\textwidth]{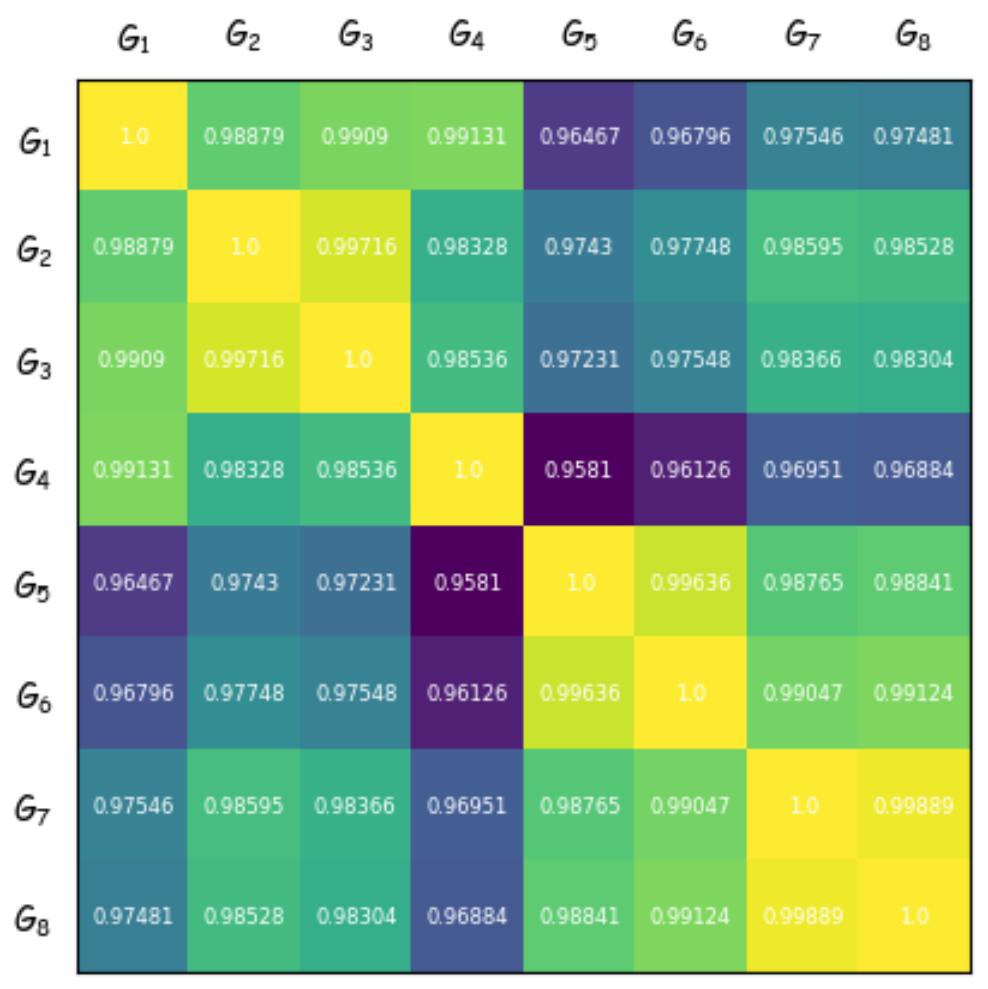} \label{fig:first}}
    \hfill
    \subfloat[Graphs with 400 nodes. Graphs $G_1$--$G_4$ correspond to $p_1=0.9$, and graphs $G_5$--$G_8$ have $p_2=0.92$.The original- all-pairs effective resistance kernel was used for the calculation.]{\includegraphics[width=0.45\textwidth]{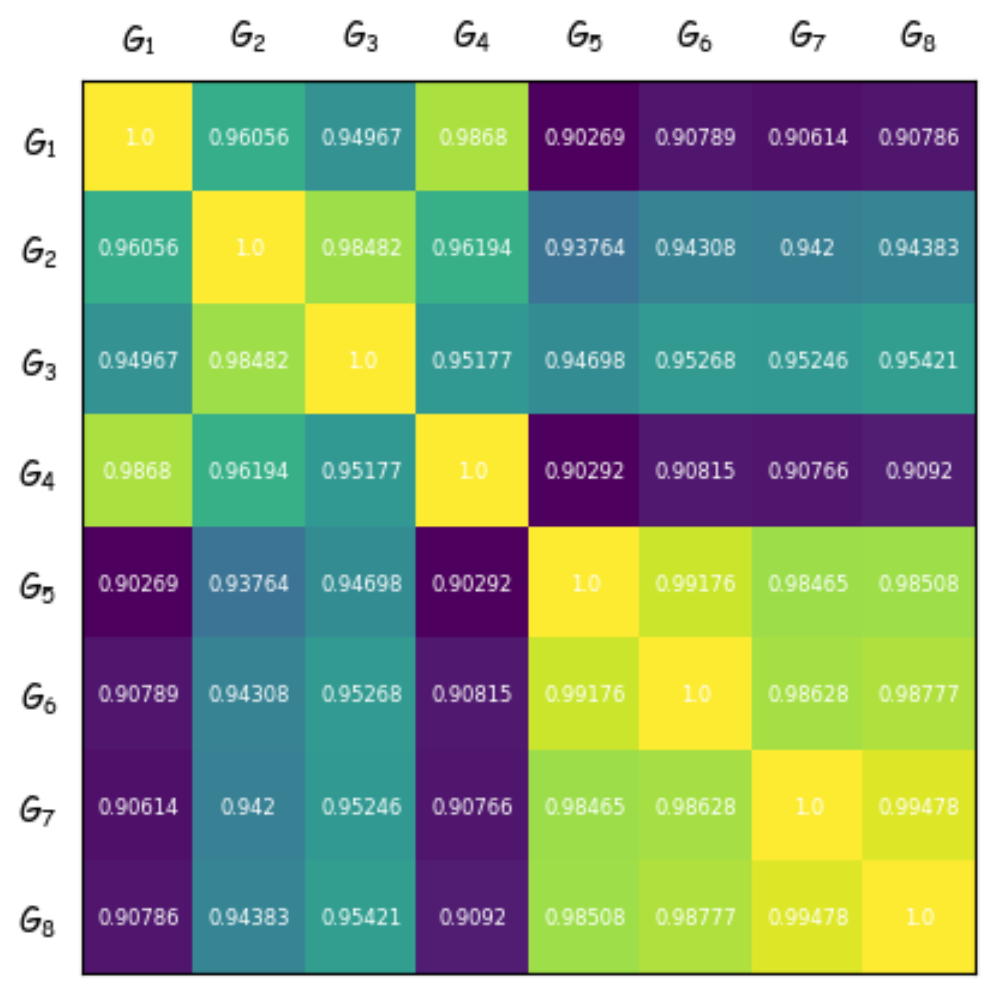} \label{fig:second}}
    \caption{Comparison of Gram matrices with kernel values of a) the reduced effective resistance kernel b) all-pairs effective resistance kernel.}
    \label{fig:comparison2}
\end{figure}

\begin{figure}[htbp]
    \centering
    \subfloat[Graphs with 600 nodes. Graphs $G_1$--$G_4$ correspond to $p_1=0.85$, and graphs $G_5$--$G_8$ have $p_2=0.86$. Reduced effective resistance kernel was used for the calculation.]{\includegraphics[width=0.45\textwidth]{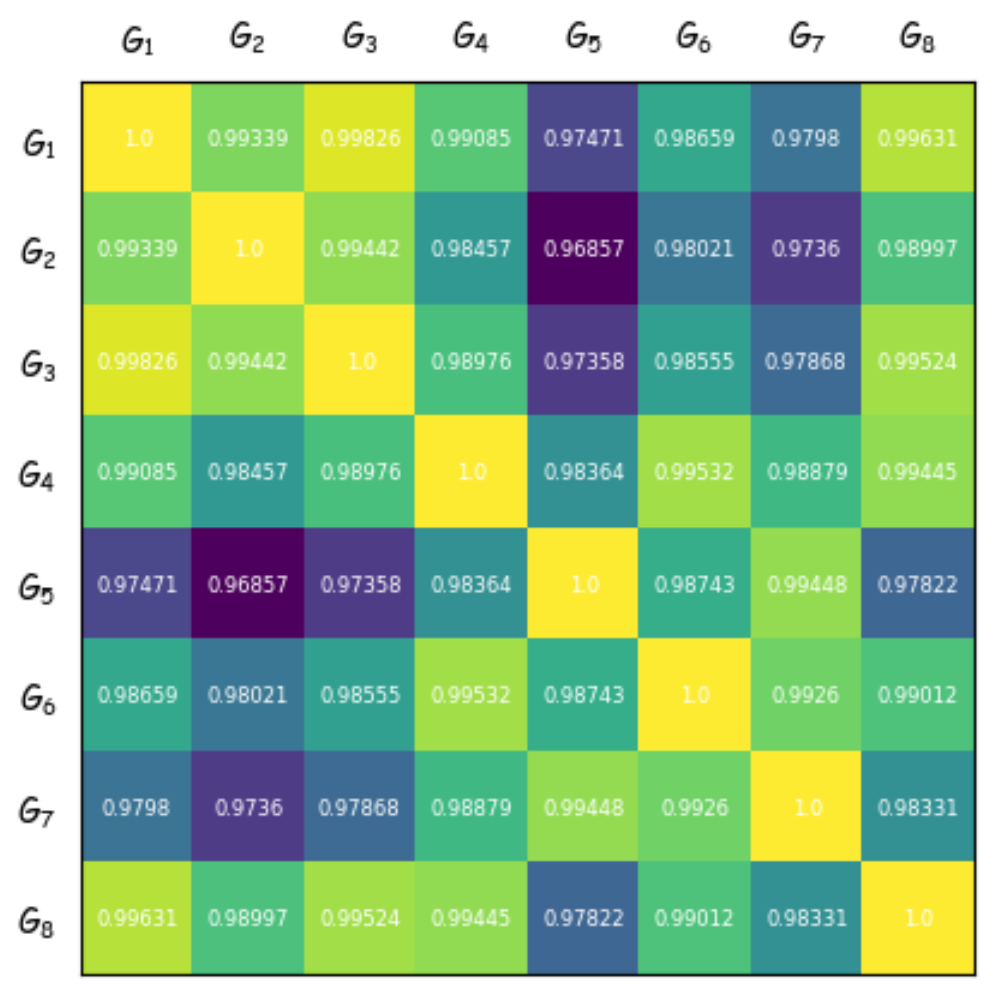} \label{fig:first}}
    \hfill
    \subfloat[Graphs with 600 nodes. Graphs $G_1$--$G_4$ correspond to $p_1=0.85$, and graphs $G_5$--$G_8$ have $p_2=0.86$. The original-all pairs effective resistance kernel was used for the calculation.]{\includegraphics[width=0.45\textwidth]{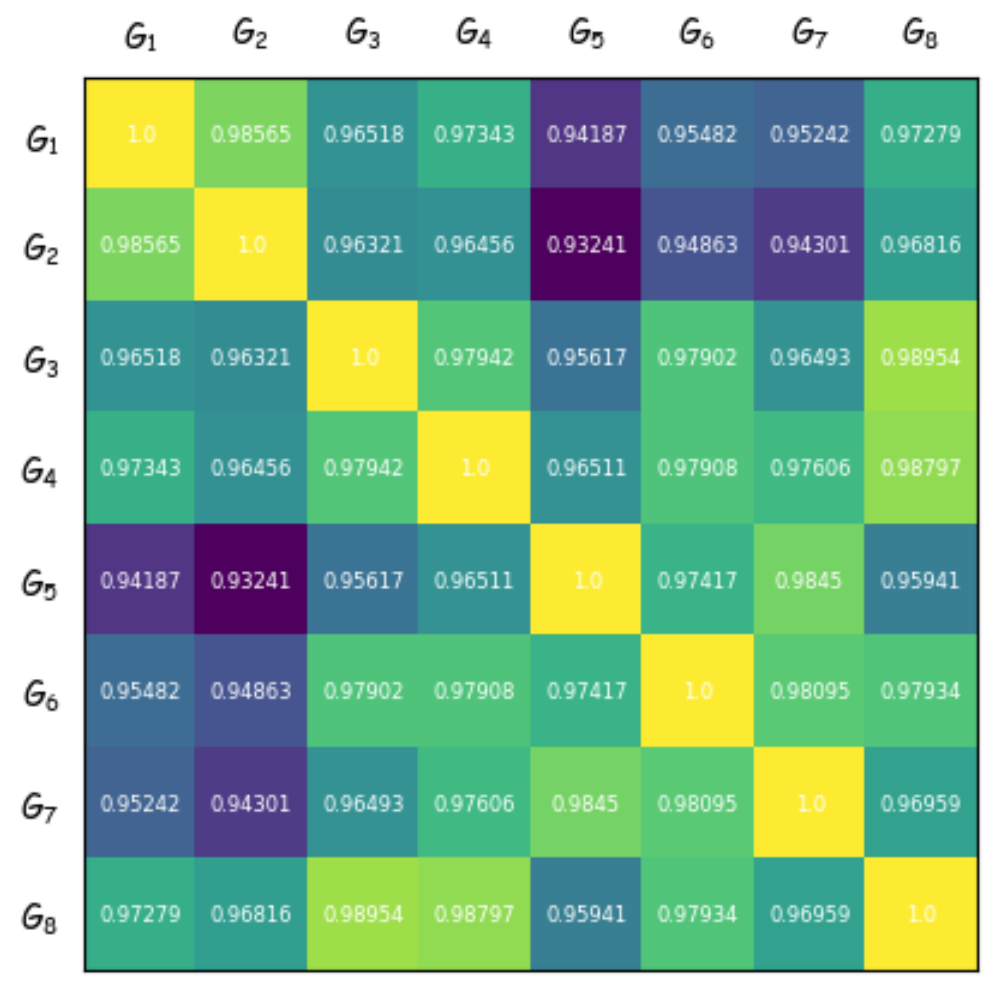} \label{fig:second}}
    \caption{Comparison of Gram matrices with kernel values of a) the reduced effective resistance kernel b) all-pairs effective resistance kernel.}
    \label{fig:comparison3}
\end{figure}
\subsection{Benchmark of effective resistance kernel on chemistry datasets}
In this section we report the accuracy of the All pairs effective resistance kernel for the standard chemistry datasets: $\tt{AIDS}$, $\tt{NCL1}$, $\tt{PTC-MR}$, $\tt{MUTAG}$, $\tt{PROTEINS}$. We compare these to the Random Walk kernel\cite{JMLR:v11:vishwanathan10a}, shortest path kernel\cite{Borgwardt2005ShortestpathKO, hermansson2015generalized}, Weisfeiler-Lehman optimal assignment kernel\cite{Shervashidze2011WeisfeilerLehmanGK} and Deep Graph Convolutional Neural Network (DGCNN)\cite{10.5555/3504035.3504579}.

\begin{center}

\begin{table}[h]
\scalebox{1.35}{
\begin{tabular}{|c|c|c|c|c|c|}
\hline
 & AIDS & NCL1 & PTC-MR & MUTAG & PROTEINS  \\
\hline
ER full  (this work) $O(n^{2.5}\sqrt{m})^*$ &$98.8\pm 0.3$ & $68.2\pm1.8$ &$61.4\pm 4.3$ & $82.8\pm 5.3$ & $65.7\pm4.2$\\ 
ER reduced (this work) $O(n\sqrt{m})^*$ & $98.1\pm 0.4$ & $66.9\pm 1.6$ & $60.4\pm 3.0$ & $82.1\pm 5.1$ & $64.2\pm 2.7$ \\
RW $O(n^3)$ & $79.7\pm 2.3$ & \tt{timeout} & $54.4\pm 9.8$ & $81.4\pm 8.9$ & $69.5\pm5.1$ \\
SP $O(n^4)$ & $99.3\pm 0.4$ & $72.5\pm2.0$ & $60.2\pm9.4$ & $82.4\pm5.5$ & $74.9\pm3.2$ \\
WL-OA $O(hm)$ & $99.2\pm 0.3$ & $86.3\pm1.6$ & $65.7\pm9.6$ & $87.0\pm5.4$ & $76.2\pm3.9$ \\
DGCNN&$99.1\pm1.4$ &$76.4\pm 1.7$ &$59.5\pm6.9$ &$84.0\pm7.1$ &$73.2\pm3.2$  \\
\hline
\end{tabular}
}
\caption{Comparison of the accuracy results (in \%) of the full and reduced effective resistance kernel (ER)-full, and -reduced, respectively, with Random Walk (RW) kernel, Shortest Path (SP) kernel,  Weisfeiler-Lehman optimal assignment kernel (WL-OA) and Deep Graph Convolutional Neural Network (DGCNN). All the results apart from the results of the kernel presented in this work are taken from\cite{nikolentzos2021graph}.}
\end{table}

\end{center}

\end{widetext}

\section{Discussion and conclusions}
In this work we explored the classification  of graphs via kernel methods and made use of the electrical network framework to assess whether the concept of effective resistance, as a distance measure on graphs, can be used to construct expressive feature vectors for graph classification tasks. Our results indicate that the reduced kernel yields comparable results to the full kernel, while requiring significantly fewer effective resistance evaluations. As illustrated in Figures (\ref{fig:comparison1}-\ref{fig:comparison3}), this computational efficiency is particularly advantageous when applied to large datasets, where other methods would be computationally prohibitive. Our time complexity analysis indicates that the reduced kernel is more efficient than most standard kernels, such as the random walk and the shortest path kernels (see table I), and has comparable complexity to the popular Weisfeiler-Lehman optimal assignment (WL-OA) kernel. 

However, despite these promising results, our benchmarks show that, while our reduced kernels have comparable performance to the random walk and shortest path kernels, it does not consistently match the accuracy of the state-of-the-art approaches, particularly the WL-OA kernel and DGCNN. As expecteed, the reduced kernel also falls behind the full ER kernel in terms of accuracy, illustrating the tradeoff between computational efficiency and classification performance. 

The tradeoff between accuracy and computational efficiency is a key consideration when selecting kernel methods. The reduced effective resistance kernel provides a substantial improvement in time complexity, lowering the cost from $O(n^2)$ in the full kernel to $O(\sqrt{n})$ which is a significant reduction, particularly for large graphs. While this remains computationally demanding for extremely large datasets, it represents a meaningful step toward scalability. Further improvements—such as achieving sub-polynomial scaling (e.g., polylogarithmic complexity)—would be necessary to make this approach fully practical for massive real-world applications.

To address these limitations, future research should aim to develop kernels that satisfy the following criteria:
\begin{itemize}
    \item low computational complexity for computing feature vectors (for individual graphs)
    \item high classification accuracy
\end{itemize}

Ideally, a kernel method should rely on a low-dimensional, yet highly expressive, feature vector that can be computed efficiently and still yield high accuracy. While the full effective resistance kernel meets the first and third criteria, it does not satisfy the second due to the computational cost of pairwise kernel evaluations. Conversely, the reduced kernel fulfills the first two criteria but lacks the accuracy required to be competitive with state-of-the-art methods. This underscores the need for further refinement of both the quantum algorithms used and, perhaps more importantly, the feature vector construction itself.

In conclusion, while the reduced effective resistance kernel offers promising computational advantages, its current form does not provide the accuracy needed to compete with state-of-the-art methods. Future work should focus on refining the kernel, exploring integration with deep learning frameworks, and expanding its applications to new domains such as finance and dynamic networks. As quantum computing progresses, these methods could become increasingly relevant, enabling efficient and accurate graph classification in real-world scenarios.



\begin{acknowledgments}
We wish to acknowledge useful discussions with Keita Kanno, who has provided us with invaluable feedback.
\end{acknowledgments}

\section{References}
\bibliography{article}

\end{document}